\documentclass[aps,preprint,floatfix,amssymb,amsmath,tightenlines,prb,endfloats]{revtex4}
\usepackage[dvips]{graphicx}
\graphicspath{{.}}
\newcommand{\setc}{\ensuremath{ \{ C \} }}

\newcommand{\tf}{\ensuremath{T_\mathrm{f}}}
\newcommand{\kb}{\ensuremath{k_{\mathrm{B}}}}

\newcommand{\fdag}{\ensuremath{F^{\dagger}}}
\newcommand{\mus}{\ensuremath{\mu\mathrm{s}}}

\newcommand{\beq}{\begin{equation}}
\newcommand{\eeq}{\end{equation}}

\makeatletter
\renewcommand\@biblabel[1]{#1.}
\makeatother

\makeatletter
\def\@cite#1#2{$(#1\if@tempswa , #2\fi)$}
\makeatother

\begin{document}


\title{\mbox{Excluded volume, local structural cooperativity,}\\
\mbox{and the polymer physics of protein folding rates}}

\author{Xianghong Qi and John J. Portman}
\affiliation{
Department of Physics, Kent State University, Kent, OH 44240
}

\date{June 1, 2007}

\begin{abstract}
A coarse-grained variational model is used to investigate the polymer
dynamics of barrier crossing for a diverse set of two-state folding
proteins.  The model gives reliable folding rate predictions provided
excluded volume terms that induce minor structural cooperativity are
included in the interaction potential.  In general, the cooperative
folding routes have sharper interfaces between folded and unfolded
regions of the folding nucleus and higher free energy barriers.  The
calculated free energy barriers are strongly correlated with native
topology as characterized by contact order.  Increasing the rigidity
of the folding nucleus changes the local structure of the transition
state ensemble non-uniformly across the set of protein studied.
Neverthless, the calculated prefactors $k_0$ are found to be
relatively uniform across the protein set, with variation in $1/k_0$
less than a factor of five.  This direct calculation justifies the
common assumption that the prefactor is roughly the same for all small
two-state folding proteins.  Using the barrier heights obtained from
the model and the best fit monomer relaxation time $30 \mathrm{ns}$,
we find that $1/k_0 \sim 1 - 5\mus$ (with average $1/k0 \sim 4\mus$).
This model can be extended to study subtle aspects of folding
such as the variation of the folding rate with stability or solvent
viscosity, and the onset of downhill folding.

\end{abstract}

\maketitle




Folding in small proteins is often well characterized as a
cooperative transition between two well-defined structural populations: an
unstructured globule ensemble and a structured folded ensemble.  The
transition rate between free energy minima is
controlled by the dynamics of passing through an unstable transition
region determined by saddlepoints in the free energy surface.
Accordingly, the rate is expected to follow Arrhenius form 
\beq
\label{eq:rate} k_f = k_0 e^{-\beta \Delta \fdag}, 
\eeq 
where $\beta = 1/\kb T$ is the inverse temperature and
$\Delta\fdag$ is the free energy difference between the unfolded and
transition state ensembles.  The exponential factor in
Eq.(\ref{eq:rate}) reflects the equilibrium population of the
transition state ensemble relative to unfolded ensemble and the
prefactor, $k_0$, is the timescale associated with the dynamics of
crossing the free energy barrier. Successful identification of specific
residues structured in the transition state ensemble by several
different theoretical models
~\cite{shoemaker:wolynes:97,shoemaker:wolynes:99a,%
portman:wolynes:98,alm:baker:99,munoz:eaton:99,%
galzitskaya:finkelstein:99,portman:wolynes:01a,portman:wolynes:01b}
and numerous simulation studies (ref.9,10,11 
and references therein) has established that the topology of the native
structure determines the folding mechanism of these proteins.  In
addition, two-state folding rates are well correlated with very simple
measures of the native state topology such as contact
order~\cite{plaxco:baker:98b,ivankov:finkelstein:03,makarov:plaxco:03,bai:zhou:04}.
While additive potentials often produce reasonable structural
characterization of the transition state ensemble, the range of
simulated folding rates and their relationship with contact order does
not agree with experiment~\cite{koga:takada:01}. In this paper we
present direct folding rate calculations that capture the trends
first noted in lattice models~\cite{jewett:plaxco:03,kaya:chan:03} and
very recently in continuum models~\cite{ejtehadi:plotkin:04,kaya:chan:05} 
leading several groups to speculate that the behavior of folding rates 
indicate enhanced structural cooperativity.

The term ``structural cooperativity'' usually refers to a mechanism by
which the presence of a structured region makes additional order more
favorable.  For example, cooperativity is greater when a contact
between two residues is more stabilized after one of the partners is
already ordered.  The additional stability is most naturally
introduced through local attractive multibody interactions associated
with coarse-grained potentials~\cite{eastwood:wolynes:01,ejtehadi:plotkin:04,wang:stell:05}.
Non-additive potentials can also be neutral with respect to disordered
and ordered residues and still increase cooperativity. For example,
destabilizing partially ordered residues near structured resides
generically sharpen the interface between folded and unfolded regions
by increasing the surface energy of the folding nucleus~\cite{wolynes:97,shoemaker:wolynes:99a}. 
Even purely repulsive interactions can enhance cooperativity.  
A good example of this is the ``induced rigidity'' (enhanced helical order) 
due to liquid crystal ordering in dense polymer solutions~\cite{kim:pincus:79,flory:matheson:84} 
and globular helical proteins~\cite{schulten:wolynes:95}.

In the present analytic model, cooperativity is introduced through
repulsive, excluded volume interactions between residues in proximity to native contact
pairs.  This potential is effectively ``neutral'' since it primarily
destabilizes partially ordered residues at the interface of the folding
nucleus.  The cooperative term of the potential is pairwise additive
in the space of all contacts, but corresponds to an effective
multi-body potential when projected onto the set of native contacts.
The particular form of cooperativity was
developed so that the calculated barrier heights remain robust with
respect to variations of excluded volume strength in the original
variational model~\cite{portman:wolynes:98,portman:wolynes:01a}.

Experimental evidence supporting a specific decomposition 
of the folding rate into the dynamic and thermodynamic factors in Eq.~\ref{eq:rate}
is necessarily indirect~\cite{yang:gruebele:03}.  While structural
predictions from models with a strong native state bias (Go-models)
are robust, the value of the barrier height (and corresponding
absolute timescale $1/k_0$) is more sensitive to details of the
model~\cite{garbuzynskiy:galzitskaya:04,henry:eaton:04}. The predicted prefactor
is commonly assumed to be roughly uniform for different proteins with
a magnitude of
$O(0.1$--$1\mus^{-1})$~\cite{yang:gruebele:03,li:thirumalai:04,naganathan:munoz:05,%
kubelka:eaton:04}, though prefactors as large as $O(100\mus^{-1})$ have
also been proposed
recently~\cite{garbuzynskiy:galzitskaya:04,plotkin:05}. While the
precise value of the prefactor is a sub-dominant  determinant
of the absolute rate, accurate estimation gives an important reference
timescale essential, for example, to identify the the fastest measured
rates as downhill (or barrier-less)
folding~\cite{yang:gruebele:03,gruebele:05,kubelka:hofrichter:05}.
Calculations for  28 two state proteins
presented in this paper predicts the prefactor is relatively uniform
on the order $O(1\mus^{-1}$), largely independent of differences in
the absolute folding rates or the native state topology.  Furthermore,
predicted folding rates agree with experimental trends provided
interaction terms favoring modest structural cooperativity are
included in the model.  In particular, the relationship between
barrier heights and contact order is 
found to be a consequence of relatively rigid folding nuclei.

\section*{Model: excluded volume and cooperativity }

The variational model developed by Portman, Takada, and
Wolynes~\cite{portman:wolynes:01a,portman:wolynes:01b} has proved
reliable in predicting the structure of the transition state ensemble
of individual proteins at the residue level of resolution%
~\cite{portman:wolynes:98,shen:wolynes:05,zong:wittung-stafshede:06}.
In this model, the free energy of partially ordered ensembles of polymer configurations
is developed through a reference
Gaussian chain inhomogeniously constrained to the native positions
by $N$ harmonic variational constraints $\setc$.
A summary of the variational model is given in the Supplemental Material.
Here, we focus on how enhanced cooperativity can be realized by the addition of
repulsive interactions between non-native contacts.

We divide the energy into two contributions
\beq\label{eq:energy}
E[\setc] =
\sum_{\mathrm{NAT}} \langle u^{\mathrm{nat}}(r_{ij}) \rangle_0 +
\sum_{\mathrm{NON-NAT}}\langle u^{\mathrm{coop}}(r_{ij}) \rangle_0
\eeq
where the subscript indicates an average over the reference Hamiltonian.
The first term represents attractive interactions
between monomers that are neighbors in the native
structure (i.e., the Go-model assumption).  
The second term, which is new to the model,
represents excluded volume interactions between non-native contact pairs.  
Before explaining the consequences of these repulsive interactions,
we first motivate the need for this contribution by considering 
the native contact potential in the original model.

The form of the interactions between native contacts is 
the sum of three Gaussians for convenience: $u^{\mathrm{nat}}(r_{ij}) =
\epsilon_{ij} \sum_{a = \mathrm{l,i,s}}\gamma_a e^{-\alpha_a
r_{ij}^2}$ where $\epsilon_{ij}$ is the strength of the
interaction~\cite{miyazawa:jernigan:96}.
Repulsive intermediate- and attractive long-range Gaussians sum to give 
a potential well with minimum $u^{\mathrm{nat}}(r_{\mathrm{min}}) = -1$ at the distance
$r_{\mathrm{min}} = 6$\AA.  The short-range Gaussian represents
excluded volume for native contact pairs; we choose $\gamma_l$ for each contact to give the
same strength at zero distance, $U(0)$.  The finite strength of the
repulsion at $r=0$ is an artifact of the potential (and the finite
native monomer density at short range), so there is some ambiguity in
determining the appropriate value for $U(0)$.  This is troubling since
it was found that the calculated barrier height is sensitive to 
the value of $U(0)$, even though the structure of the transition state
ensemble is relatively robust for most proteins.

The sensitivity of $\Delta \fdag/\kb \tf$ on the excluded volume
strength $U(0)$ indicates that the cooperativity in the model is
relatively low.  This can be understood by considering the
short-distance pair density of a partially ordered chain,
$n_{ij}(\mathbf{r}) = \langle \delta(\mathbf{r} -
\mathbf{r}_{ij})\rangle_0$. Integrating over angles gives the radial
pair density 
\beq 
n_{ij}(r)
\sim \frac{1}{\sqrt{a^2 \delta G_{ij}}} \frac{r}{s_{ij}} 
\sinh \left( \frac{3rs_{ij}}{a^2\delta G_{ij} } \right)
\exp \left( - \frac{3}{2} \frac{ r^2 + s^2_{ij} }{a^2\delta G_{ij}}\right)
\eeq
where the correlations $G_{ij} = \langle \delta \mathbf{r}_i \cdot \delta \mathbf{r}_j \rangle_0/a^2$ 
and  $\delta G_{ij} = G_{ii} + G_{jj} - 2G_{ij}$ is the magnitude of
the fluctuations about the relative mean separation $\mathbf{s}_{ij}= \sum (G_{ij}-G_{jk}) C_k \mathbf{r}_{k}^{\mathrm{N}}$;and 
$a = 3.8$\AA is the distance between adjacent monomers.
The weight at short distances ($r < r_0$, $r_0$ is  excluded volume interaction length scale)
is small when the pair is sufficiently delocalized ($ r_0 \ll a \sqrt{\delta G_{ij}}$) or
sufficiently localized ($a \sqrt{\delta G_{ij}} \ll s_{ij}$).
Consequently, the sensitivity of the barrier height on the short
distance repulsion is due to the partially ordered residues
in the transition state ensemble.  Increasing the cooperativity by destabilizing
partially ordered residues makes the barrier height less sensitive to $U(0)$.
To this end, we modify the interactions
between native contacts through a repulsive potential  between
residues in close proximity to native contacts
$u^{\mathrm{coop}}(r_{ij})/\epsilon_0 = U(0) \exp\left[-\alpha_\mathrm{s}
r_{ij}^2\right]$.
This term increases cooperativity of partially structured ensembles
by encouraging residues surrounding a native ordered pair to either be
delocalized or ordered themselves to reduce short-distance density overlap.

\begin{figure}[p]
\centering
\includegraphics[width=2.5in]{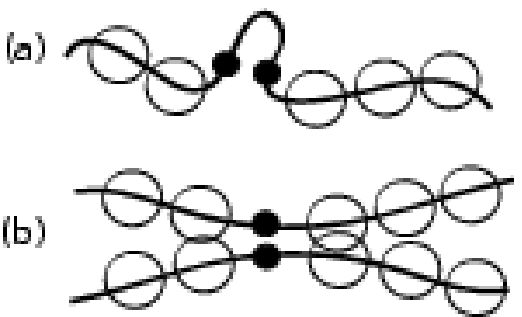}
\caption{Cooperativity of local and non-local contacts.(a)Localization of a 
short-ranged pair induces little cooperativity
because the monomer density of the surrounding residues is not significantly
altered by a partially formed local contact . (b)In contrast, if the contact pair is
non-local sequence, this bring together larger regions residues (of
the order of the persistence length) in proximity, increasing the
cooperativity .
The dependence of the cooperativity on sequence length 
is reminiscent of cooperative desolvation between
folded segments~\cite{lum:weeks:99,li:chan:2005}.
\label{fig:coop}}
\end{figure}

When viewed as an effective potential involving only native contact pairs,
the repulsion between non-native contacts effectively induces local multi-bodied interactions.
Due to chain connectivity, structured regions that are sufficiently 
non-local in sequence have greater cooperativity 
(see Fig.
~\ref{fig:coop}). 
Accordingly, we define a reduced set of non-native contacts:
for every native pair $(i,j)$ with $|i-j| \ge 12$ we include pairs within a
window $[i\pm 4, j \pm 4]$ and eliminate duplicates or native contact
pairs from the sum.
With this convention, 
the barrier height varies less than about 1--2$\kb \tf$,
over a wide range of $U(0)$ $(5 \le U(0) \le 60)$.  In
the following we take $U(0) = 50$.  

The parameters of the model are the same as given in 
\citealp[Ref.][]{portman:wolynes:01a},
except: (i) the magnitude $U(0)$ is fixed for each contact; (ii) cooperativity term;
(iii) the radius of gyration of the globule is set by the chain length according to the scaling
law given in 
\citealp[Ref.][]{jha:sosnick:05}.
We note that with these parameters, the folding route for $\lambda$-repressor studied in 
\citealp[Ref.][]{portman:wolynes:01a},
is structurally similar the cooperative folding route,
though the barrier is approximately two times larger.

\section*{Results}

\begin{figure}[p]
\centering
\includegraphics[width=3.0in]{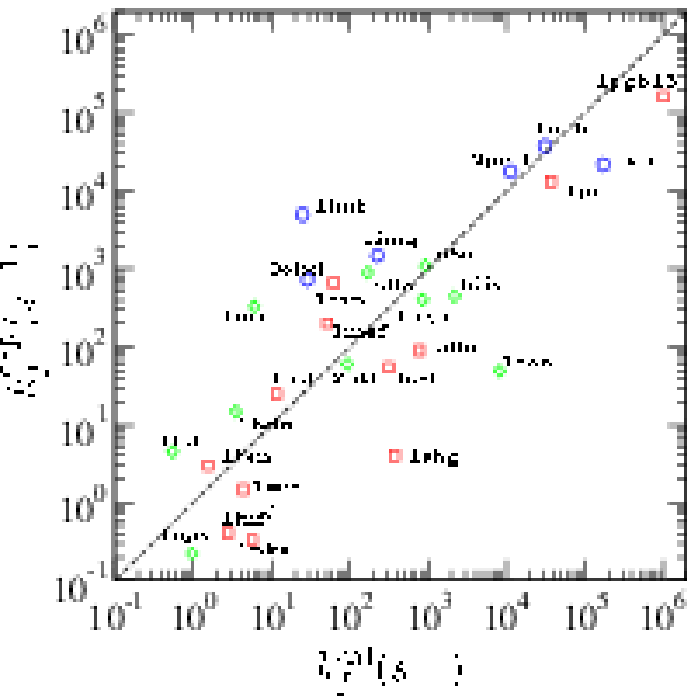}
\caption{(color) Comparison between experimental and calculated folding rates.
The correlation coefficient is $r=0.8$, with $p$-value $p = 4.4\times 10^{-7}$.  
(circle denotes $\alpha-$protein, square denotes $\beta-$protein, and 
diamond denotes $\alpha\beta-$protein.). Kinetic data for proteins 1enh, 1vii, 1hdn  
were taken from \citealp[Ref.][]{bai:zhou:04}; kinetic data for the rest of proteins 
were taken from \citealp[Ref.][]{ivankov:finkelstein:03}.Here, x-ray structured proteins 
are: 1pgb,1coa,1csp,1c8c,1pin,1lmb,1enh, 1fkb,1urn,1shg,1ten,$^*$1div,$^*$1fnf$^9$,$^*$ 1pgb16; NMR structured 
proteins are: 1o6x,1imq,1srl,2ptl,1aps,2abd,$^*$1a0n,$^*$2pdd,$^*$1mef,$^*$1psf, $^*$1wit, $^*$1pks,$^*$1vii,$^*$1hdn.
(Proteins not used in $\phi$-value analysis are indicated by $^*$ )
\label{fig:kf}}
\end{figure}

\textbf{Folding rates and prefactors.~} 
We calculated the prefactors and folding routes of 28
two state folding proteins.  The
corresponding folding rates at the transition midpoint, 
$k_f = k_0 e^{-\Delta\fdag/\kb \tf}$,
are shown in Fig.~\ref{fig:kf}. 
Calculated rates in absolute units depend
on the timescale set by the monomer relaxation rate
$\sigma = 3D_0/a^2$ which we take as a fitting parameter.  As shown in
Fig.~\ref{fig:kf}, the predicted and measured rates are well
correlated ($r = 0.8$) with agreement within an order of magnitude for
80\% of the proteins. 

The best fit monomer relaxation time $1/\sigma = 30 \mathrm{ns}$
is on the order of the timescale of unfolding a helical segment~\cite{ivankov:finkelstein:01}.
With this microscopic timescale, the longest
relaxation time of a chain of 100 monomers is approximately $\tau_{R} \sim O(10\mus)$
which compares well~\cite{degennes:85} with the timescale for the fastest collapse kinetics
measured in proteins and polypeptides~\cite{shastry:roder:98,qiu:hagen:03,magg:schmid:04}.
On the other hand, $1/\sigma$ is an order of
magnitude slower than estimates obtained from an effective diffusion
coefficient inferred from loop closure experiments of small peptides ($\sim 1$ns) and two
orders of magnitude slower than estimates from bare diffusion
coefficients of the monomer ($\sim 100$ ps)~\cite{lapidus:hofrichter:99}.
The source of small effective diffusion coefficients associated with
simple Gaussian models is not fully understood~\cite{lapidus:hofrichter:02,yeh:hummer:02,portman:03}.
Nevertheless, results from recent experiments on small peptides under different
solvent conditions indicate that intrachain interactions 
(that can be interpreted kinetically as a kind of internal friction) induce local 
activation barriers that renormalize the effective monomer
diffusion coefficient~\cite{buscaglia:hofrichter:06,moglich:kiefhaber:06}.
Although controversial, internal friction
may explain why the speed limit of protein folding is fixed at 
$\sim O(0.1$--$1.0\mus)$~\cite{qiu:hagen:04}.

\begin{figure}[p]
\centering
\includegraphics[width=3.0in]{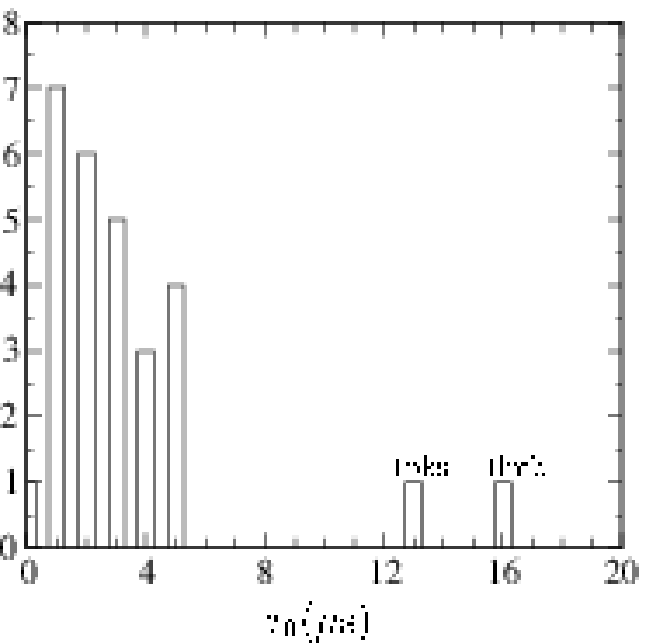}
\caption{Histogram of the inverse folding prefactor $\tau_0 = 1/k_0$ 
for the 28 two-state proteins. Ordinate is number of proteins.
\label{fig:tau0}}
\end{figure}

Even though the prefactor of each protein is calculated individually,
its value in absolute units ultimately depend on the calculated
barrier heights through the microscopic timescale $\sigma$.  Relative
prefactors, on the other hand, are independent of this fitting
parameter.  As shown in Fig.~\ref{fig:tau0}, the distribution of
prefactors is relatively uniform, varying within a factor of 5 for most
proteins. 
Using the fitted value for $\sigma$, $\tau_0 =1/k_0$ varies mainly
between $1 \mus$ to $5 \mus $ with an average $\bar{\tau}_0 = 4 \mus$.
Given this narrow distribution, it is not surprising that a uniform
prefactor of $\bar{k}_0 = (4 \mus)^{-1}$ gives essentially the same
correlation to the measured and predicted rates (data not shown).
Thus, direct calculation of the barrier crossing dynamics gives solid
evidence supporting the common assumption that the folding rate
prefactor is largely independent of topology.  Recent work by Henry
and Eaton also suggests the prefactor
is relatively uniform across two-state folding
proteins based on analysis of folding rates from a
different set of analytic models~\cite{henry:eaton:04}.  
The value for the average prefactor
$\bar{k}_0 \sim 10^5 s^{-1}$ agrees within an order of magnitude with
estimates based on semi-empirical and theoretical
models~\cite{munoz:eaton:99,alm:baker:02,kouza:thirumalai:06} as well
as analysis of thermodynamic data from differential scanning
calorimetry~\cite{naganathan:munoz:06}.  This value also is consistent
with the fastest measured rates $\sim 1\mus$, if the timescale for downhill
folding is approximated by the Arrhenius rate with a vanishing
barrier~\cite{yang:gruebele:03,kubelka:hofrichter:05}.

A closer look at the two proteins (1lmb and 1pks) with exceptionally
small calculated prefactors reveals that in each case the unstable
mode becomes degenerate at a stability near the transition
midpoint. The structure of the transition ensemble changes sharply,
though continuously, as the temperature crosses the degenerate point.
In particular, the curvature of the unstable mode (and consequently
the calculated prefactor) sharply vanishes in a cusp
catastrophe~\cite{gilmore:81}.  Away from these isolated temperatures,
the prefactors return to the range exhibited by the majority of the
proteins studied. Several of the proteins studied have similar rapid
changes of the transition state as a function of temperature, occurring
at temperature sufficiently far away from the midpoint so that the
prefactor is relatively unaffected near $\tf$.  In this high dimensional
model, catastrophes can be
generally expected as local minima and saddle-points merge at isolated
values of the control parameters (e.g., temperature).  
The shape of the calculated prefactor versus
temperature is thus determined by these degenerate points.  For
example, even for a route with a single transition state, the
meta-stable unfolded or folded minimum disappear in a fold catastrophe
at the limit of stability (spinodal) for both low and high
temperatures~\cite{wales:01,bogdan:wales:04}. If there are no other
catastrophes, the calculated prefactor obtains a maximum at an
intermediate temperature and vanishes at the spinodals.  Near the
maximum the prefactor varies much more slowly with temperature than
near the spinodal. This generic shape of the prefactor is interesting
since it can account qualitatively for non-linear dependence of the
rate with stability (chevron turnovers) and may
indicate kinetic signatures anticipating the onset of downhill
folding.

Nevertheless, interpreting these results requires some care.  The
harmonic expansion of the free energy is not expected to accurately
reflect the global curvature of the free energy over $\sim \fdag \pm
\kb T$ when the local curvature is very small.  For these cases, it is
likely that the formalism should be modified away from strictly local
curvatures to get accurate estimates of the prefactor.  Even if the
renormalized prefactor is found to be relatively constant, the rapid
change of the order parameter at the transition state that accompanies
a catastrophe may itself account for cheveron roll-over, similar to the
transition state switching mechanism suggested by
Oliveberg~\cite{otzen:oliveberg:99}. The subtle
variation of the prefactor and free energy barrier height with
stability is an important issue that has yet to be thoroughly
explored.\\

\begin{figure*}[p]
\centering
\includegraphics[width=6.5in]{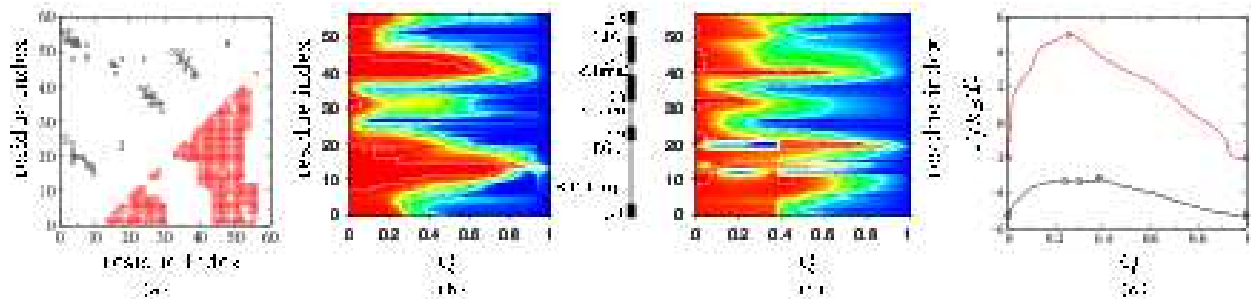}
\caption{(color) Folding route for protein Src tyrosine kinase SH3 domain(PDB 1srl)
characterized locally by the normalized native density
$\tilde{\rho}_i =
(\rho_i -\rho_i(G))/(\rho_i(N)
-\rho_i(G))$ and the global progress coordinate
$Q = 1/N\sum \tilde{\rho}_i$.In the center is a diagram of the protein secondary structure with 
five $\beta-$strand $\beta-$barrel ,three loops(RT,n-Src,distal),a diverging turn and a $3_{10}$ helix.
(a) Contact map:  black squares are native contacts, and  red squares are non-native pairs.
The local and global structure along the folding route with cooperativity:
(b) and without cooperativity (c).The degree of structural 
localization of each residue is reflected in the colors, linearly scaled from
red ($\tilde{\rho}_i = 0$) to blue ($\tilde{\rho}_i = 1$). 
(d) Free energy profile as a function of $Q$: the red curve is folding route with
cooperativity, and the black curve is without cooperativity. The circles denote the 
critical points defining the folding route, while the curves are steepest descents paths. 
Experimental $\phi$-values suggest that 
the $\beta2-\beta3-\beta4$ are ordered in the transition state ensemble~\cite{riddle:baker:99}
in agreement the structure predicted  from the model.
\label{fig:1srl} }
\end{figure*}

\textbf{Free energy profiles and folding routes.~} 
In the context of 
identifying kinetic trends for all two state proteins, a complete theory of the
folding mechanism must reliably predict structural properties 
of the transition state ensemble in addition to absolute folding rates. 
The formation of local order along the folding route can be characterized
by the degree of localization about the native positions 
\beq \label{eq:natdens}
\rho_i[\setc] = \left\langle \exp\left[
-3\alpha^{\mathrm{N}}(\mathbf{r}_i - \mathbf{r}_i^{\mathrm{N}})^2/2a^2
\right] \right\rangle_0.
\eeq 
We refer to $\rho_i$ as the native density.
Comparing the folding profiles and structural localization of the residues
shown in Fig.~\ref{fig:1srl}  illustrates the cooperative nature of the folding
routes induced by the repulsive non-native interactions.  While the coarse
grained structures of the transition state ensembles are similar for
this protein, the residues order more gradually in the non-cooperative
routes. Still, even for the cooperative route, the interface has a finite width as
the structural ensembles retains some partial ordering of the residues.
The sharper interface of the cooperative route is also
accompanied by a significantly larger barrier.

\begin{figure*}[p]
\centering
\includegraphics[width=6.0in]{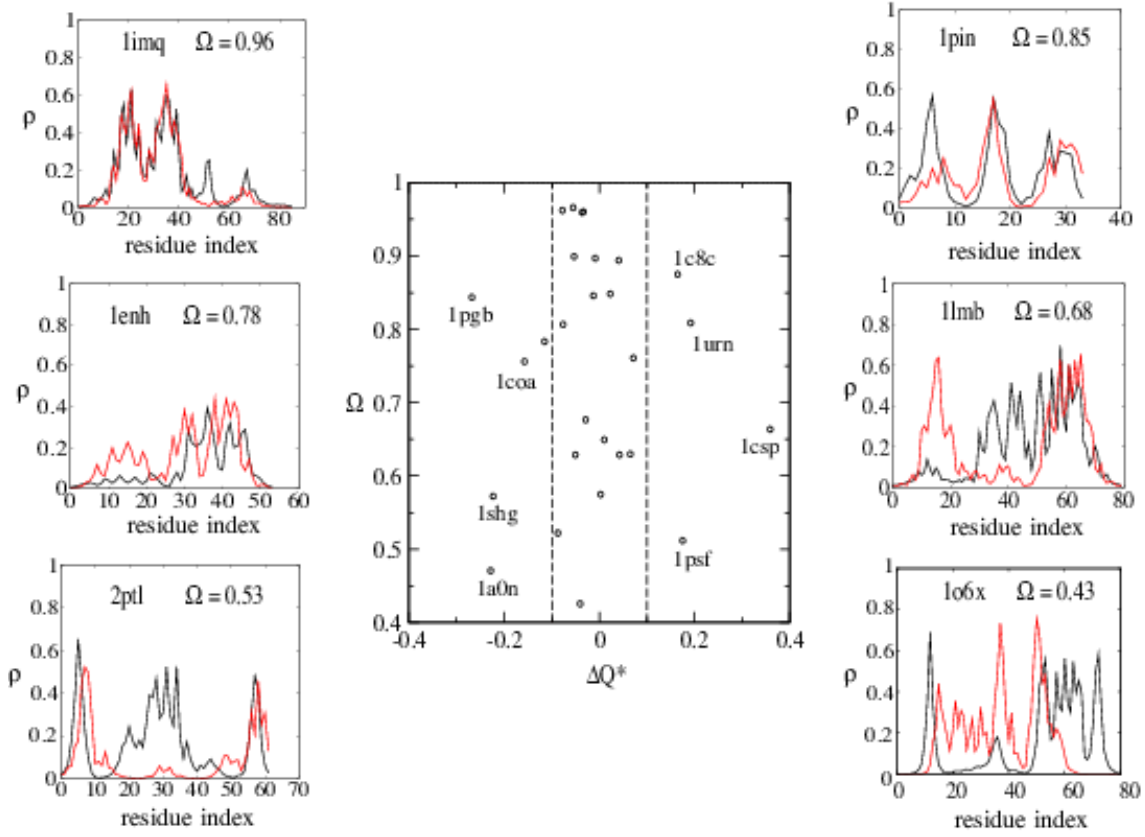}
\caption{(color) Comparision of the structure of cooperative and non-cooperative transition state ensembles.
(Center)The change in the local structure of the transition state ensemble characterized
by the cross-correlation coefficient
$\Omega$ plotted against the change in the global order parameter $\Delta Q^{\star}$ . 
(Left and right) Typical examples of the native density profile for
various values of overlap $\Omega$. In each example, the red line corresponds to the cooperative route,
and, the black line corresponds to the non-cooperative route. The corresponding correlation coefficient between
measured and predicted phi-value are 1pin:0.77; 1lmb:0.84;1enh:0.87; 2ptl:0.0;1imq:-0.1;1o6x:-0.2.
\label{fig:overlap}}
\end{figure*}

The effect of cooperativity on the structure of the transition state
ensemble is complicated to describe in general.  Cooperativity narrows
the interface by destabilizing partially ordered residues in favor of
either more ordered or more disordered.  Whether a particular
interfacial residue is excluded or incorporated into the folding
nucleus is a subtle question, determined by the delicate balance
between changes in entropy and energy due to localization. One measure
to characterize changes in local structural order is the cross
correlation 
\beq\label{eq:overlap}
\Omega=\hat{\mathbf{\rho}}_{coop}\cdot\hat{\mathbf{\rho}}_{non-\\coop},
\eeq 
where $\hat{\mathbf{\rho}}_{coop}$ or
$\hat{\mathbf{\rho}}_{non-coop}$ denote unit vectors with elements
$\mathbf{\rho}_i[\setc]$ for transition state ensembles with or
without cooperativity, respectively. Fig.~\ref{fig:overlap} shows the
value $\Omega$ for each protein as well as a typical example of the
overlap of native densities evaluated at $Q^{\star}$.  For $80\%$
proteins studied, the overlap between the transition state ensemble
structures is greater than $60\%$. Nevertheless, the variation of $\Omega$
indicates this form of cooperativity does not effect every protein
uniformly.

Changes of the transition state ensemble can be also be characterized by
the variation of the global order parameter $\Delta Q^{\star}$.  As
shown in Fig.~\ref{fig:overlap}, the majority of the proteins studied
have $|\Delta Q^{\star}| \le 0.1$.  In terms of global order, the
$\alpha$-helical proteins are not very sensitive to cooperativity, though the
local structure of the transition state ensemble can change
significantly.  For $\beta$ and $\alpha/\beta$ proteins, some
systematic errors in the calculated barrier height can be associated
with relatively large changes in the global order.
For proteins with $\Delta Q^{\star} > 0.1$ (1urn, 1c8c, 1psf, 1csp),
the model overestimates the barrier heights, while for proteins with
$\Delta Q^{\star} < 0.1$ (1pgb, 1a0n, 1coa, 1shg) the model
underestimates the barrier height. (See Fig.~\ref{fig:kf}. ) This
trend may be particular to form of 
cooperativity used in this model.

Direct comparison between theoretical and measured $\phi$-values
shows that cooperative routes generally have significantly higher
correlation with experiment.  Following Garbuzynskiy \textit{et al.}~\cite{garbuzynskiy:galzitskaya:04}, 
we make a distinction between contact maps
obtained from native structures determined by x-ray crystallography and
those from the first model of an NMR structure or minimized averaged NMR structure.  
Overall, the theory predicts $\phi$ values for studied x-ray structures
reasonably well. Still, there are exceptions.  Of the 11 x-ray structures
(see caption in Fig.~\ref{fig:kf}), two proteins( 1shg,  1ten)  have large negative correlations.
The average correlation coefficient 
for nine remaining proteins increases from 0.33 (for non-cooperative routes)
to 0.6 (cooperative routes).  Predictions of $\phi$-values 
for NMR determined structures
is significantly worse with the average 0.1 for both non-cooperative routes and cooperative routes.
In Fig.~\ref{fig:overlap} , we give three examples for x-ray structures and NMR structures respectively.\\

\begin{figure}[p]
\centering
\includegraphics[width=3.0in]{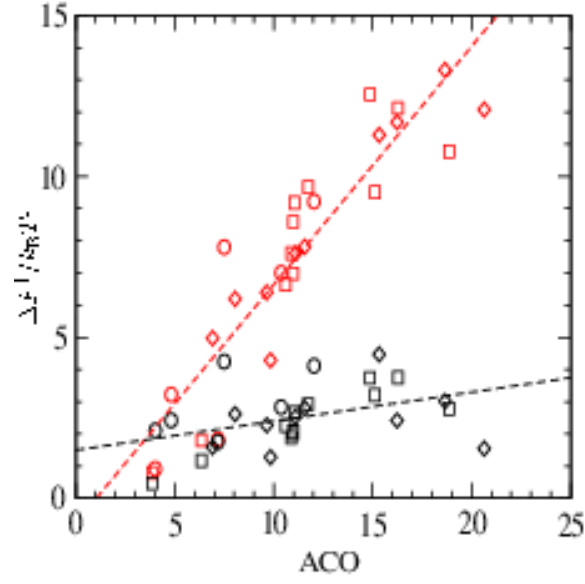}
\caption{(color) Free energy barrier $\fdag/\kb \tf$ plotted against the absolute contact order,
$ACO=1/N_{\mathrm{con}}\sum|i-j|$ where the sum is over the 
$N_{\mathrm{con}}$ native contacts pairs. 
Red points correspond to cooperative folding routes, and black points correspond to non-cooperative
routes. Symbols have the same meaning as Fig.~\ref{fig:kf}.
For the cooperative routes, the correlation coefficient is
$r= 0.91$ with p-value $p = 1.6*10^{-11}$; for non-cooperative routes, the
correlation coefficient if $r = 0.41$ with p-value $p = 0.03$.
\label{fig:DFT_ACO}}
\end{figure}

\textbf{Folding barriers and absolute contact order.~} Since the
prefactors are relatively uniform, the wide variation of relative
folding rates is determined by differences in free energy barriers.
To investigate the relationship between barrier heights and native
topology, we consider the correlation between the free energy barriers
and the absolute contact order~\cite{ivankov:finkelstein:03}.
Fig.~\ref{fig:DFT_ACO} shows that the calculated barrier height is
highly correlated $(r = 0.9)$ with absolute contact order when the
cooperativity term included in the model.  The barrier heights
calculated without cooperativity does not show significant correlation
with absolute contact order. This indicates that the relationship
between native topology and the folding rate is sensitive to the
rigidity of the folding nucleus.  This may in fact be a robust
result, largely independent of the details of reasonable potentials
that increase local cooperativity between native
contacts~\cite{ejtehadi:plotkin:04}.

Assuming the prefactor is roughly uniform, the range of measured rates
for this protein set correspond a range of free energy barrier heights of
about $\sim 14\kb \tf$, in agreement with the calculated barriers.
In contrast, the range of barriers for the non-cooperative routes 
spans only $\sim 5\kb\tf$. Interestingly, this is the same range determined through
coarse-grained Go-model
simulations~\cite{koga:takada:01,chavez:clementi:04,stefan:chan:06}.
Furthermore, the low correlation between contact order and barrier heights
of non-cooperative routes is also reminiscent
of results from Go-model simulations~\cite{koga:takada:01}. Together,
these results suggest that the cooperativity of typical Go-model
simulations based on two-body pair potential is too
low~\cite{jewett:plaxco:03,kaya:chan:03,ejtehadi:plotkin:04,kaya:chan:05}.

\section{Conclusion}
The repulsive potential between residues in proximity to native
contacts is a convenient way to alleviate sensitivity on the excluded
volume strength in the original model.  This was successful 
because the potential enhances cooperativity of the model.
Our point of view is that the nature
of the interface of the folding nucleus is key in
determining the behavior of folding rates and mechanisms, regardless
of the specific form of cooperative interactions or the microscopic
origins.  If the qualitative results from this study can be extended
beyond this variation model, it is likely to be limited to models that 
enhance cooperativity locally.  Because these results are robust with
respect to the excluded volume strength $U(0)$, 
the model lacks flexibility to explore a wide range
of surface tensions.  It will be very interesting to see if these conclusions hold when the
interfacial surface tension is controlled directly through, for
example, the formalism of density functional theory of first
order nucleation. 

\begin{acknowledgments}
This work was supported in part by grant awarded by the Ohio Board of Regents
Research Challenge program.
\end{acknowledgments}

\small 



\end{document}






\textbf{\Large{Supporting Text}}\\  \\ \\


In this supplement, we outline the variational model developed in 
ref.1 and ref.2.

\textbf{Variational Model.~}
A configuration of the
protein is modeled by the $N$ position vectors of the $\alpha$ carbons
of the polypeptide backbone.  Partially ordered ensembles of polymer
configurations are described by a reference Hamiltonian corresponding
to a harmonic chain inhomogeneously constrained to the native structure
$\{\mathbf{r}^{\mathrm{N}}_i \}$
\beq 
\beta \mathcal{H}_0 = 
\frac{3}{2a^2} \sum_{ij} \mathbf{r}_i\cdot \Gamma^{0}_{ij} \cdot \mathbf{r}_j 
+ \sum_{i} C_i ( \mathbf{r}_i - \mathbf{r}^{\mathrm{N}}_i )^2.  
\eeq
The first term enforces polymeric constraints,
with $\Gamma^{(0)}_{ij}$ determined by the correlations of a freely rotating 
chain~\cite{bixon:zwanzig:78}. 
The values of the 
harmonic constraints, $\setc$, control the magnitude of the
fluctuations of each monomer about the native structure 
(i.e., the temperature factors).
Ensembles of partially ordered configurations are
represented by monomer densities $n_i(\mathbf{r}) =
\langle \delta(\mathbf{r} - \mathbf{r}_i)\rangle_0$ described as
Gaussian distributions 
with variance $G_{ii} = \langle |\delta \mathbf{r}_i|^2 \rangle_0/a^2$
about the mean position of the ith monomer,
$\mathbf{s}_i = \sum G_{ij} C_j \mathbf{r}^{\mathrm{N}}_i$.
Here, the correlations $G_{ij} = \langle \delta \mathbf{r}_i \cdot \delta \mathbf{r}_j \rangle_0/a^2$
depend on both the polymeric and structural constraints
through $G^{-1}_{ij} = \Gamma_{ij} + C_i\delta_{ij}$.

The population of a partially ordered ensemble specified by the constraints
$\setc$ is controlled by the free energy $F[\{C\}] = E[\{C\}] - TS[\{C\}]$.
Here, $T$ is the temperature, $S[\{C\}]$ is the entropy loss due to
localizing the residues around mean positions, and $E[\{C\}]$ is the energy associated
with the partially ordered ensemble.
The values of the variational constraints are determined by the critical points in
the free energy surface. For each local minimum or saddlepoint, there corresponds to 
a set of $N$ variational constraints that solve $\partial_{C_i} F[\setc] = 0$.
Transition states ensembles are identified as the saddle-points
of $F[\setc]$ that connect local minima (in a steepest descents sense). A 
folding route is the series of minimum-saddlepoint-minimum which connect
the globule and native minima of $F[\setc]$.\\

\textbf{Barrier Crossing Dynamics.~} 
The formation of local order along the folding route is characterized
by the degree of localization about the native positions 
\beq \label{eq:natdens}
\rho_i[\setc] = \left\langle \exp\left[
-3\alpha^{\mathrm{N}}(\mathbf{r}_i - \mathbf{r}_i^{\mathrm{N}})^2/2a^2
\right] \right\rangle_0.
\eeq 
We refer to $\rho_i$ as the native density.
The prefactor, in turn, is determined by the
growth rate of $\rho_i(t)$ 
along the unstable mode of the free energy. In the formalism developed in
\citealp[ref.][]{portman:wolynes:01b}, 
the growth rate is developed
through the polymer dynamics of the constrained chain.  Since the
reference chain is harmonic, the correlation function between monomers $i$ and $j$,
 $G_{ij}(t) =
\langle \mathbf{r}_i(t)\cdot \mathbf{r}_j(0) \rangle_0$, can be expressed as
a sum over normal (Rouse) modes 
\beq 
G_{ij}(t) = \sum_p \frac{Q_{ip}Q_{jp}}{\lambda_p} e^{-\sigma\lambda_p t} 
\eeq 
where the coefficients $Q_{ip}$ and relaxation rates ${\lambda_p}$ are
determined by 
\beq 
(\Gamma^0_{ij} + C_i^\star\delta_{ij})Q_{jp} = \lambda_p Q_{ip}.  
\eeq 
Here, the monomer relaxation rate $\sigma
= 3D_0/a^2$ is set by the effective bond length $a$ and monomer
diffusion coefficient $D_0$.  

The effective diffusion matrix
corresponding to the native density dynamics, $\mu_{ij}(t)$,
can be defined through
the Laplace transform of the correlation functions $\mathcal{C}_{ij}(t) =
\langle \rho_i(t) \rho_j(0)\rangle_0 -
\langle \rho_i\rangle_0 \langle
\rho_j\rangle_0$: 
\beq 
\hat{\mu}(\omega) = \mathcal{C}(0)\cdot\hat{\mathcal{C}}^{-1}(\omega) \cdot \mathcal{C}(0) 
-\omega \mathcal{C}(0).
\eeq 
Here, $\mathcal{C}(t)$ is determined by the polymer dynamics through the monomer correlations $G(t)$.  
Since $\mu_{ij}(t)$ depends on both the structure of the constrained ensemble as well as time, this formalism
gives a microscopic realization of the effective diffusion coefficient used in more general formulations
of landscape theory~\cite{bryngelson:wolynes:89,lee:wang:03}.

Finally, the
prefactor is given by $k_0 = |\omega|/2\pi$, where the growth rate
$\omega$ is the negative eigenvalue of \beq \hat{\mu}_{ij}(|\omega|)
\Gamma_{jk} u_k = -|\omega| u_{ij} \eeq where 
$\Gamma_{ij} =
\partial^2\beta F[\{C^{\star}\}]/\partial\rho_i\partial\rho_j$ 
is the curvature of the free energy with respect to the native density
evaluated at the saddlepoint.
